\documentclass[3p]{elsarticle}
\usepackage{chemformula}
\usepackage{lineno,hyperref}

\journal{Spectrochimica Acta Part A: Molecular and Biomolecular}

\begin{document}

\begin{frontmatter}

\title{Vacuum Ultraviolet Photoabsorption Spectra of Icy Isoprene and its Oligomers}

\author[1]{R Ramachandran}
\author[2]{S Pavithraa}
\author[1]{J K Meka}
\author[1]{ K K Rahul}
\author[3,4]{J -I Lo}
\author[3]{S -L Chou}
\author[4]{B -M Cheng}
\author[5]{B N Rajasekhar}
\author[1]{Anil Bhardwaj}
\author[6]{N J Mason}
\author[1]{B Sivaraman\corref{mycorrespondingauthor}}
\cortext[mycorrespondingauthor]{Corresponding author}
\ead{bhala@prl.res.in}

\address[1]{Physical Research Laboratory, India.}
\address[2]{Department of Applied Chemistry and Institute of Molecular Sciences, National Chiao Tung University, Hsinchu, Taiwan.}
\address[3]{National Synchrotron Radiation Research Centre, Hsinchu, Taiwan.}
\address[4]{Department of Medical Research, Hualien Tzu Chi Hospital, Buddhist Tzu Chi Medical Foundation, Hualien, Taiwan.}
\address[5]{Atomic and Molecular Physics Division, Bhabha Atomic Research Center, Mumbai, India.}
\address[6]{School of Physical Sciences, University of Kent, Canterbury, UK.}

\begin{abstract}
	Isoprene and its oligomers, terpenes, are expected to be present, along with other complex organic molecules, in the diverse environments of the ISM and in the solar system. Due to insufficient spectral information of these molecules at low temperature, the detection and understanding of the importance of these molecules has been rather incomplete. For this purpose, we have carried out the vacuum ultraviolet (VUV) photoabsorption measurements on pure molecular ices of isoprene and a few simple terpenes: limonene, $\alpha$-pinene and $\beta$-pinene by forming icy mantles on cold dust analogues. From these experiments, we report the first low temperature (10 K) VUV spectra of isoprene and its oligomers limonene, $\alpha$-pinene and $\beta$-pinene. VUV photoabsorption spectra of all the molecules reported here reveal similarities in the ice and gas phase as expected, with an exception of isoprene, where a prominent red shift is observed in the ice phase absorption. This unqiue property of isoprene along with distinctive absorption at longer wavelengths supports its candidature for detection on icy bodies. \\
	\\ 
\end{abstract}


\end{frontmatter}


\section{Introduction}
\par Isoprene is a major biogenic volatile organic compound emitted on Earth (500 Tg C year\textsuperscript{-1} in similar quantities to methane) \cite{mcgenity2018}. It accounts for $\approx$40\% of the biogenic volatile organic compounds emitted by plants \cite{hillier2019} and also a principal hydrocarbon exhaled by humans \cite{gelmont1981}. Isoprene (C\textsubscript{5}H\textsubscript{8}) is an unsaturated hydrocarbon and readily reacts with \ch{^{.}OH}, O\textsubscript{2}, NO\textsubscript{x}, etc., hence it has a short lifetime in the terrestrial atmosphere ($<$3 hours) \cite{liu2018}. Multiple isoprene units together form the terpenes, which are naturally occurring compounds and belong to the biggest class of secondary metabolites, which are responsible for the characteristic aroma of various plants, including cannabis. Natural rubber is also a polymer of isoprene, indeed isoprene was first synthesised by burning the natural rubber \cite{faraday1826}. Even though the isoprene is short lived, the stable isoprenoids (terpenes with an additional functional group) are well identified in fossils \cite{summons1988} \cite{brooks1969} \cite{nwandie2018}, some of which formed even before the origin of atmospheric oxygen \cite{nwandie2018}. It is also hypothesized that isoprene can be central to understanding the origin of membranes and hence abiogenesis \cite{segre2001}. \\
\par Astrophysically, isoprene has been suggested to be a possible biological marker for the presence of life in other planets and exoplanets \cite{zhan2021}.  G{\"u}nay et al., \cite{gunay2018} synthesized isoprene based interstellar dust analogs in the laboratory and suggested isoprene to be a good candidate to be a component of the interstellar dust based on the comparison of the aliphatic absorption feature of isoprene dust analogues with that of the observed ISM absorption spectra. Presently, isoprene and terpenes have been rather elusive in astronomical observations despite more complex organic molecules being discovered in the astrophysical environments. However, with high resolution James Webb Space Telescope (JWST) in the pipeline, there is a higher probability of discovering them in the near future. Hence it is necessary to have a detailed spectral profile of these molecules under such astrophysical conditions to aid their discovery in space. In this paper, we present the VUV spectral profile of ices of isoprene, a monocyclic monoterpene: limonene and the bicyclic monoterpene isomers: $\alpha$-pinene and $\beta$-pinene, collected over a range of astrophysical temperatures. The infrared spectra of these molecular ices will be presented and discussed in a future publication.

\section{Methodology}
A series of experiments to measure VUV spectra of isoprene, limonene, $\alpha$-pinene and $\beta$-pinene in the ice phase were carried out at the National Synchrotron Radiation Research Centre (NSRRC), Taiwan. The VUV light, from beamline BL-03 of the Taiwan Light Source (TLS), was dispersed with a cylindrical grating monochromator (focal length 6 m) on a bending magnet beamline of a storage ring (1.5 GeV) and passed through a gold mesh with transmission about 90\%. To monitor and normalize the beam, the photocurrent of the gold mesh was detected with an electrometer (Keithley 6512). The molecules were deposited onto a LiF window maintained at 10 K using a closed cycle helium system. The absorption spectra in the VUV region are measured to the limit of transmission of optical components, which in this case was about 106 nm. \\

For the absorption measurements, the VUV light from the synchrotron was perpendicular to the solid samples, and the transmitted light passed through a glass window coated with sodium salicylate to convert the incident UV light to visible light, which was measured with a photomultiplier tube (Hamamatsu R943-02) operating in a photon-counting mode. Detailed discussions of the experimental setup, liquid sample preparation, and spectral acquisition have been presented in an earlier publication \cite{lu2008}. All the samples were of high purity (from Sigma Aldrich) and were deposited onto a LiF substrate held at 10 K. Immediately after deposition, a spectrum was recorded. The sample was then heated to higher temperatures, at the rate of 5 K min\textsuperscript{-1} and a spectrum was recorded at every 20 K steps to observe any variations in the VUV spectral signatures.

\section{Results and Discussion}

\par
\begin{figure}[!h]
	\centering
	\includegraphics[width=7cm,height=5.5cm]{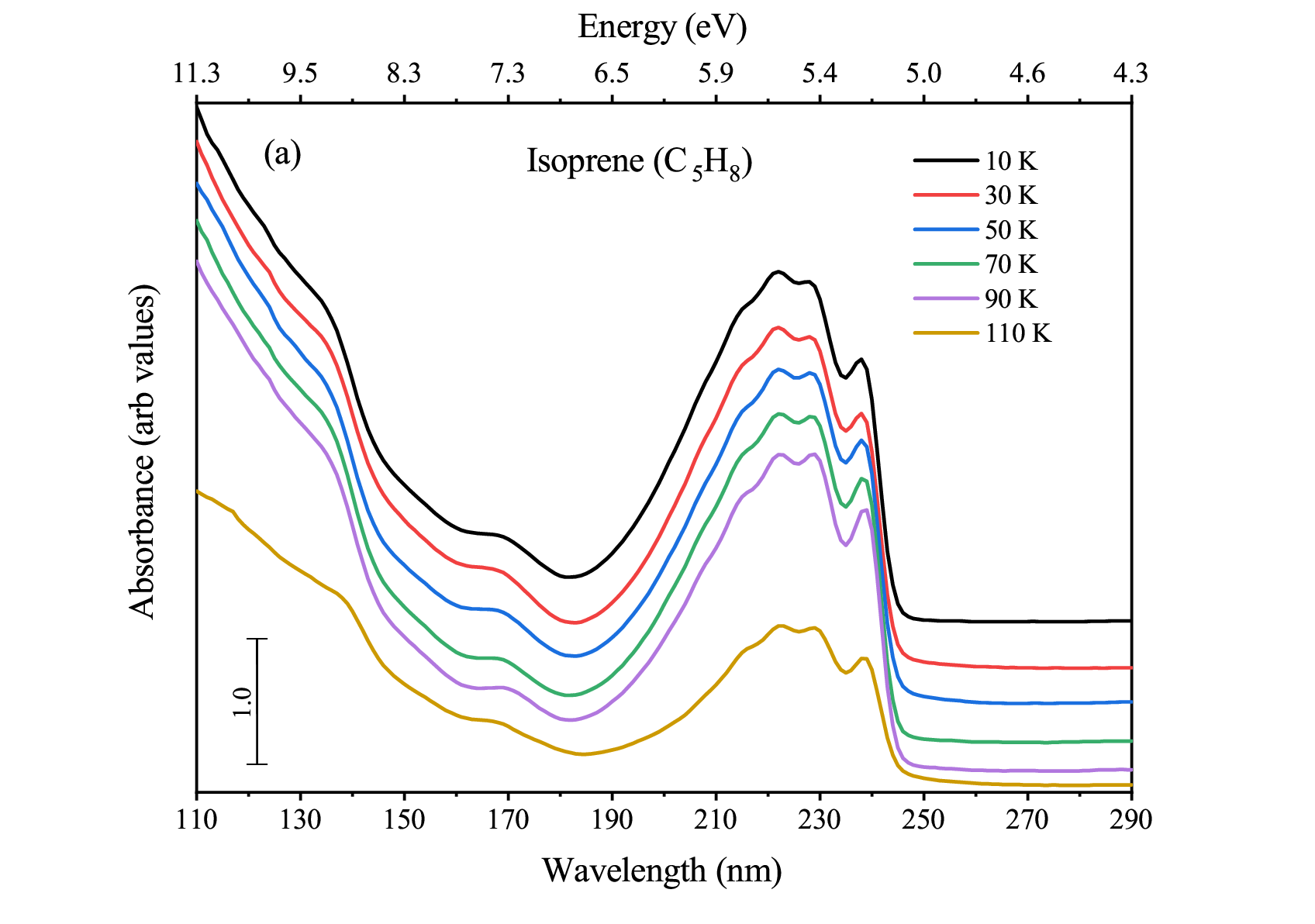}
	\includegraphics[width=7cm,height=5.5cm]{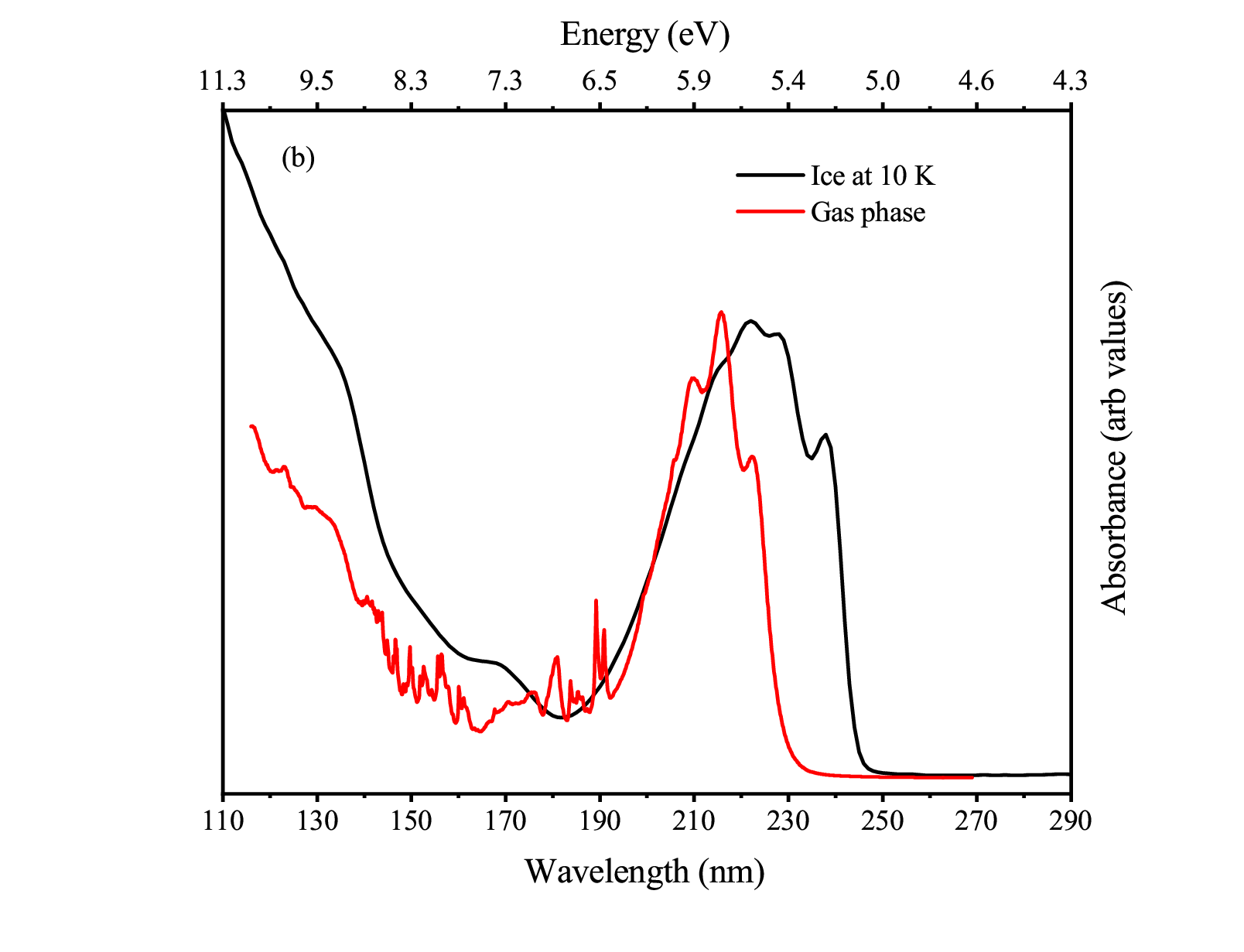}	
	\caption{ (a) Temperature-dependent VUV photoabsorption spectrum of isoprene ices. (b) Comparison of VUV spectra of isoprene between ice phase at 10 K and gas phase (from \cite{martin2001}).}
	\label{fig1}
\end{figure}

 Figure \ref{fig1}(a) shows a VUV spectrum of isoprene ice recorded over the range 290 nm - 110  nm (4.27 eV - 11.27 eV) at 10 K. The isoprene ice was found to show absorption from  $\sim$250 nm (4.96 eV) to 110 nm (11.27 eV). From 250 nm (4.96 eV) to 180 nm (6.89 eV), an intense broad absorption band is observed with several peaks at 238 nm (5.21 eV), 228 nm (5.44 eV), 222 nm (5.59 eV) and $\sim$214 nm (5.77 eV). From 180 nm (6.89 eV) to 110 nm (11.27 eV) an absorption slope from low to the higher wavelengths was observed with two peaks at 170 nm (7.29 eV) and 135 nm (9.18 eV). Above 250 nm (4.96 eV), there was no absorption. Upon warming the sample from 10 K to higher temperatures, at the rate of 5 K min\textsuperscript{-1}, the spectra remained unchanged. However, a significant drop in the absorption intensity is noticed for the spectrum recorded at 110 K. With further increase in temperature the molecule was found to desorb from the substrate. Though there are no significant changes observed in the temperature-dependent spectra of isoprene, we expect a phase change from amorphous to crystalline ice to have taken place while warming the ice as isoprene is a simple molecule among the terpenes. In the case of isoprene ice, the VUV spectral signatures are insensitive to the phase change; therefore, infrared spectral studies are required, which can clearly show the phase changes in the isoprene ice. \\

\par Figure \ref{fig1}(b) compares the gas phase spectrum \cite{martin2001} with that of the solid isoprene. The spectrum of ice phase isoprene is significantly different from that of the gas phase, in particular there is a complete absence of Rydberg states in the solid phase spectrum. The region between 5 eV and 9 eV in the gas phase consists of 3 bands which are assigned to (1$\pi$* $\leftarrow$ 2$\pi$), (3p$\pi$/2$\pi$*(C–C) $\leftarrow$ 2$\pi$) and (2$\pi$* $\leftarrow$ 2$\pi$) transitions \cite{martin2001}. Most fine structures present in the gas phase, due to (3p$\pi$/2$\pi$*(C–C) $\leftarrow$ 2$\pi$) Rydberg transition, were quenched and therefore were absent in the ice phase spectrum \cite{mason2006} \cite{bhala2012} \cite{bhuin2014}. The intense absorption band at higher wavelength was observed to be much broader with a significant red shift in the absorption for isoprene in the ice phase in comparison with the gas phase spectrum.\\
\begin{figure}[!h]
	\centering
	\includegraphics[width=7cm,height=5.5cm]{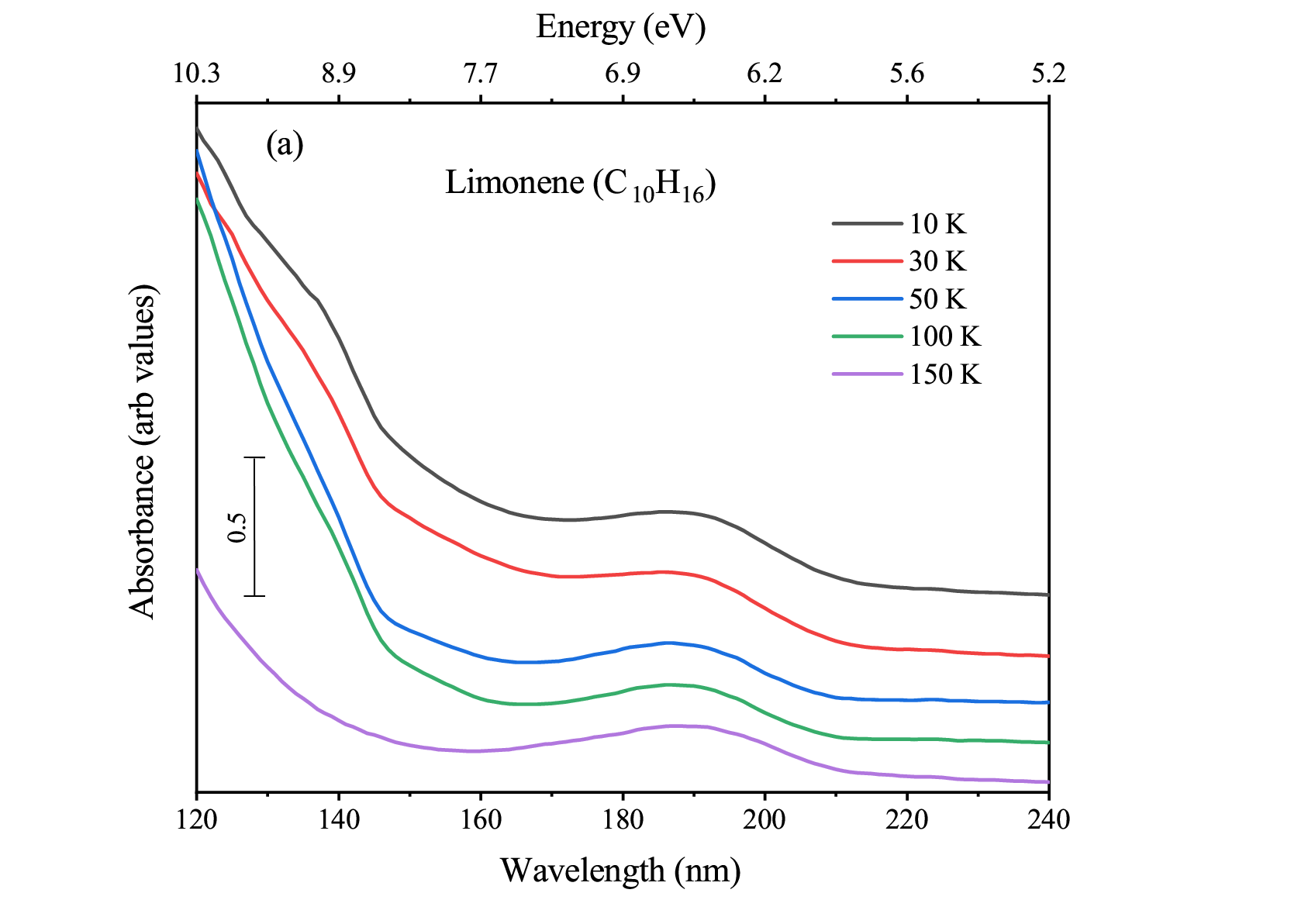}
	\includegraphics[width=7cm,height=5.5cm]{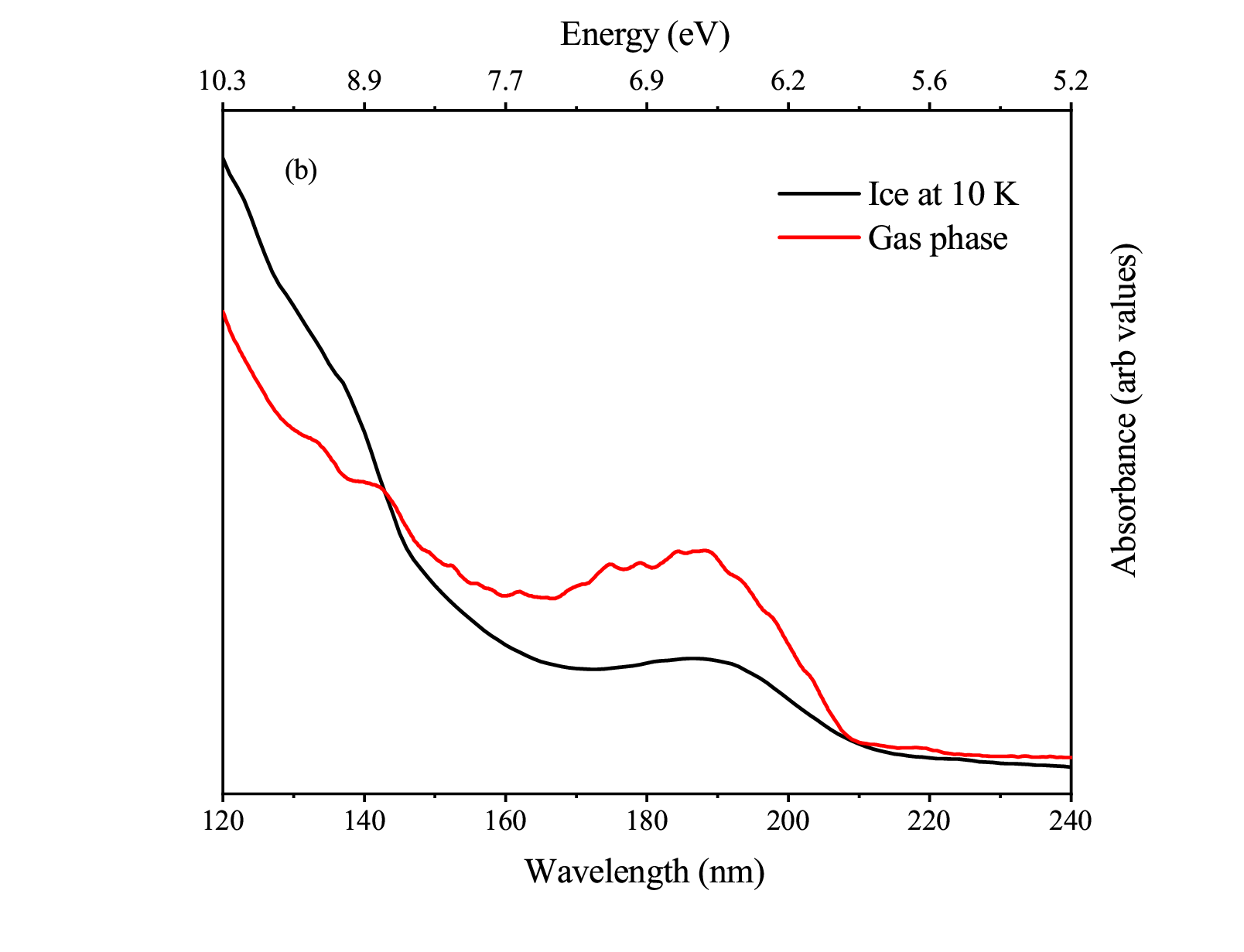}
	\caption{ (a) Temperature-dependent VUV photoabsorption spectrum of limonene ices. (b) Comparison of VUV spectra of limonene between ice phase at 10 K and gas phase (from \cite{smialek2011}). }
	\label{fig2}
\end{figure}

\par The VUV photoabsorption spectrum of limonene ice deposited at 10 K , shown in Figure \ref{fig2}(a) is observed to have absorption from 240 nm - 120 nm (5.20 eV - 10.33 eV); a broad absorption from 220 nm (5.64 eV) to 160 nm (7.75 eV) with a peak centering around 190 nm (6.52 eV) and an absorption slope from 160 nm - 110 nm (containing a peak centering $\sim$138 nm (8.98 eV)). Upon warming the ice to higher temperatures, at the rate of 5 K min\textsuperscript{-1}, no significant changes were observed in the spectrum up until 150 K where the absorption intensity reduced, indicating the onset of limonene sublimation. The VUV spectrum of ice phase limonene at 10 K is compared with the gas phase spectrum recorded \cite{smialek2011} over the same energy range (Figure \ref{fig2}(b)). We observe that in the region 5 eV - 8.5 eV, all the Rydberg states present in the gas phase spectrum are absent in the ice phase, which has rather smooth bands lacking any noticeable structures. The intensity of the band centered at 6.5 eV, which was hypothesized to be purely due to Rydberg states \cite{smialek2011}, had a reduced intensity in the ice phase.\\

\begin{figure}[!h]
	\centering
	\includegraphics[width=7cm,height=5.5cm]{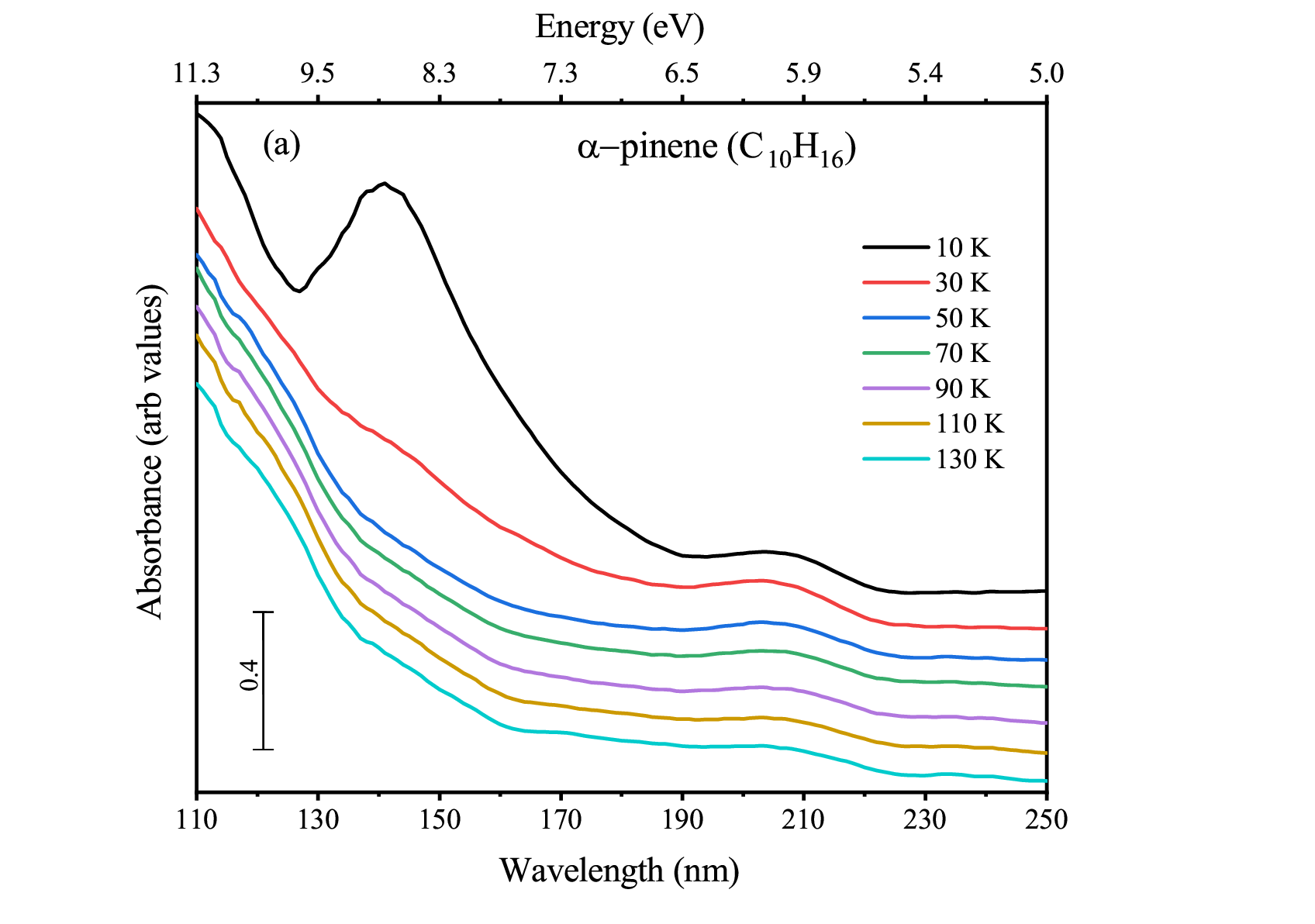}
	\includegraphics[width=7cm,height=5.5cm]{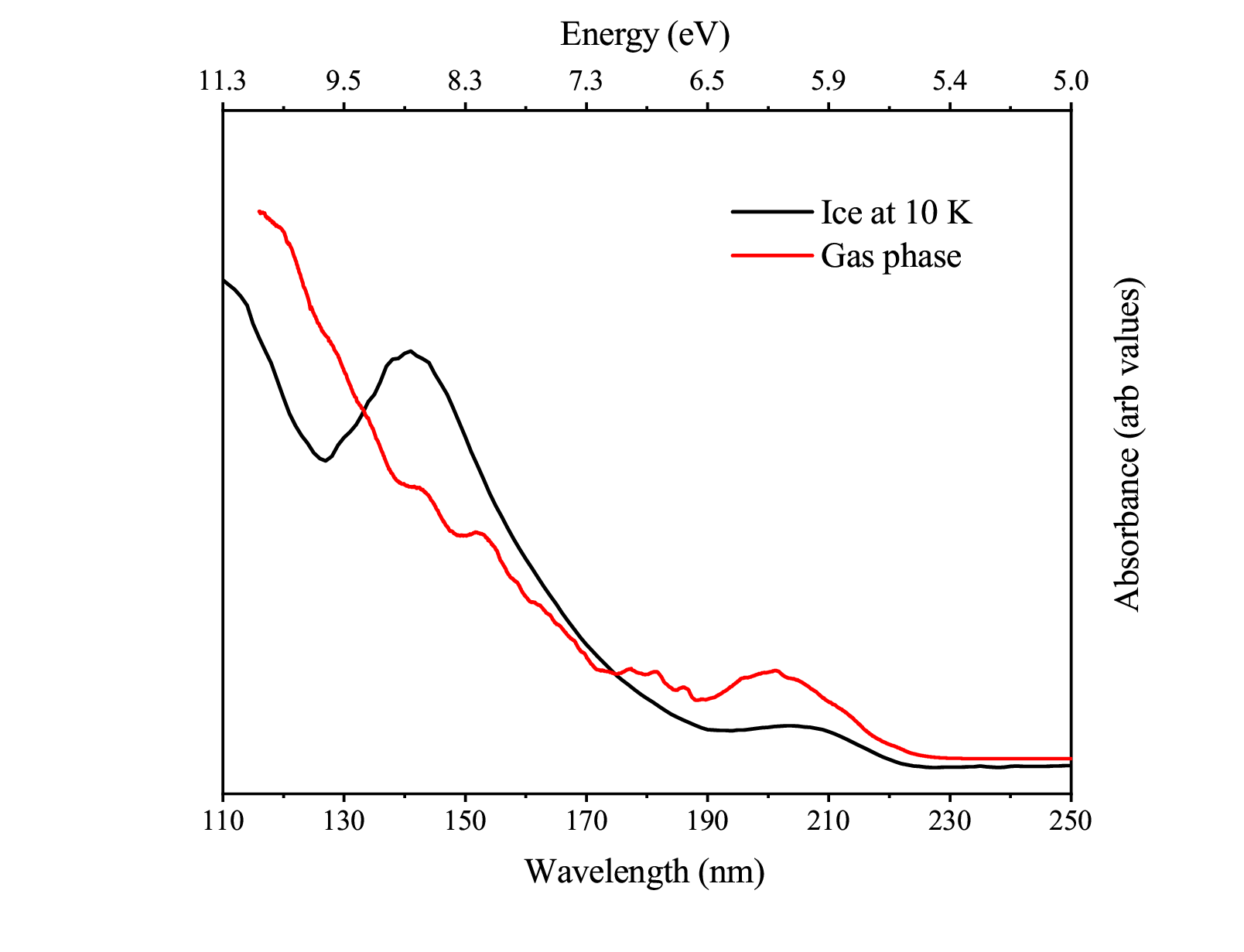}
	\caption{ (a) Temperature-dependent VUV photoabsorption spectrum of $\alpha$-pinene ices. (b) Comparison of VUV spectra of $\alpha$-pinene between ice phase at 10 K and gas phase (from \cite{kubala2009}). }
	\label{fig3}
\end{figure}

\par The VUV spectrum of the $\alpha$-pinene (Figure \ref{fig3}(a)) ice deposited at 10 K in the range 250 nm - 110 nm (4.96 eV - 11.27 eV) has two broad bands, one between 220 nm - 190 nm (5.64 eV - 6.52 eV) and the other between 190 nm - 125 nm (6.52 eV - 9.92 eV) peaking at around  205 nm (6.05 eV) and 145 nm (8.55 eV), respectively. While the first band has a strong absorption, the other band is weak. On heating the $\alpha$-pinene ice to 30 K, the strong absorption feature at 220 nm - 190 nm (5.64 eV - 6.52 eV) completely disappears, and beyond that temperature, the spectral features remain unchanged. This could be due to the highly porous amorphous ice deposited at 10 K compacting to a less porous amorphous ice when applying little thermal energy. Upon warming the ice to higher temperatures, up to 130 K, no significant changes were observed in the spectra. \\

\par
\begin{figure}[!h]
	\centering	
	\includegraphics[width=7cm,height=5.5cm]{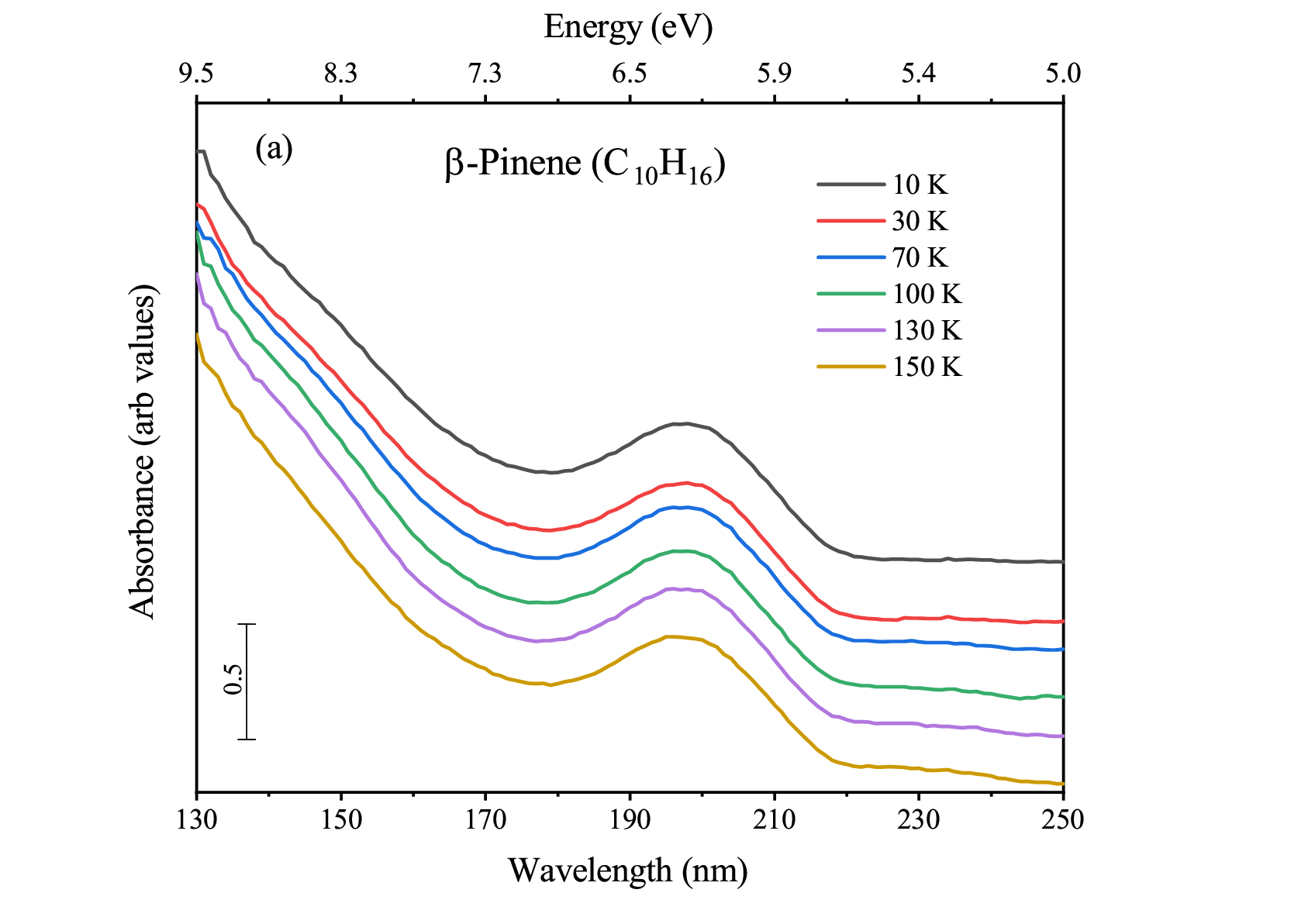}
	\includegraphics[width=7cm,height=5.5cm]{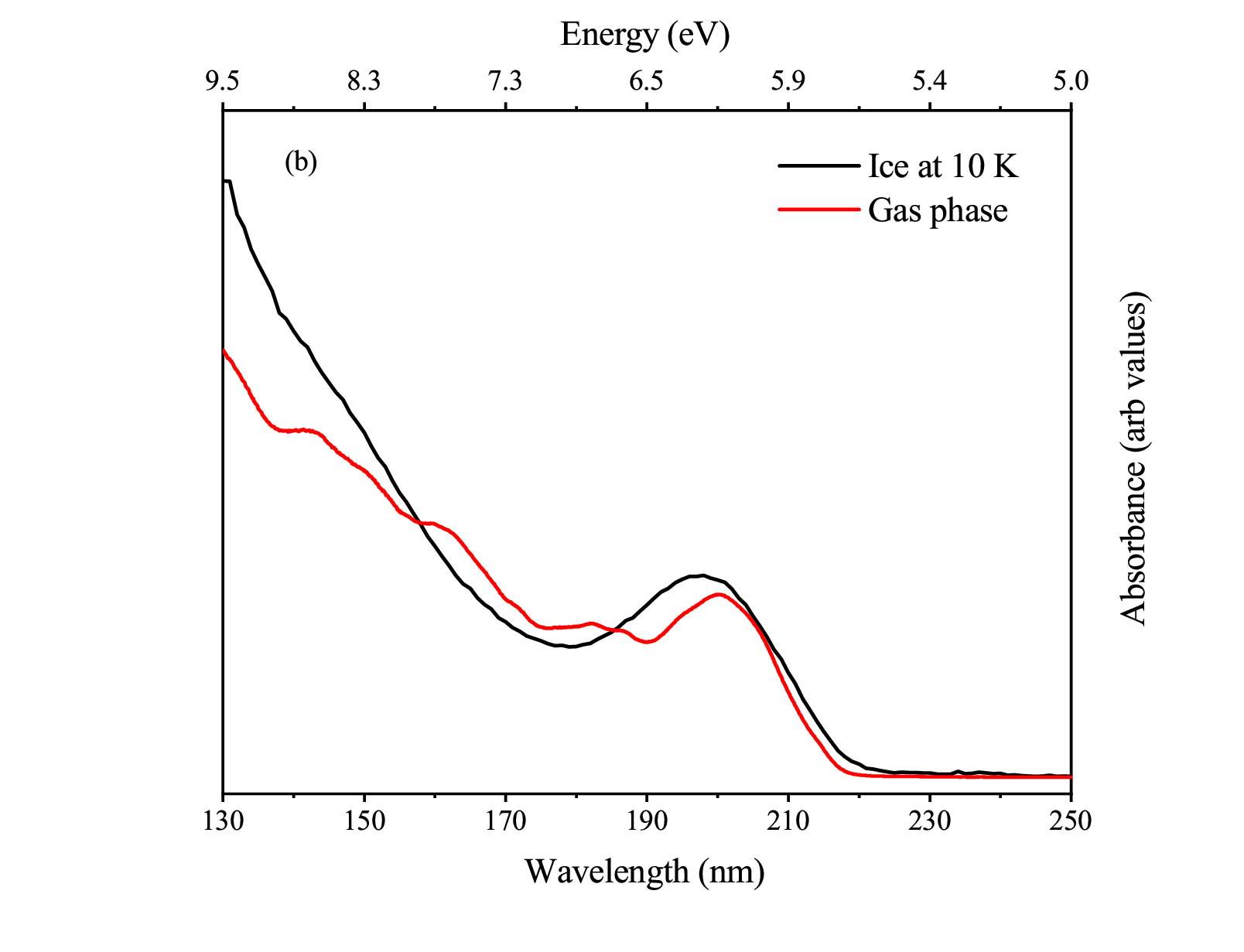}
	\caption{ (a) Temperature dependent vacuum ultraviolet photoabsorption spectrum of $\beta$-pinene ices. (b) Comparison of VUV spectra of $\beta$-pinene between ice phase at 10 K and gas phase (from \cite{kubala2009}). }
	\label{fig4}
\end{figure}

 For $\beta$-pinene deposited at 10 K (Figure \ref{fig4}(a)), the VUV spectrum in the 250 nm - 130 nm (4.96 eV - 9.54 eV) region is found to have only one band in the range 220 nm - 175 nm (5.64 eV - 7.08 eV) peaking at 200 nm (6.2 eV). Even after heating up to 150 K, the spectral features remain unchanged. While the gas phase spectra of $\alpha$-pinene and $\beta$-pinene (Figure \ref{fig3}(b) \& \ref{fig4}(b) respectively) beyond 7 eV is monotonically increasing, it is not the case in the ice phase. As mentioned before, there appears a broad absorption band, peeking at 145 nm (8.55 eV), for $\alpha$-pinene at 10 K which is missing in the gas phase (and also for ices above 30 K). The absorption band at the longer wavelength, 220 nm - 190 nm (5.64 eV - 6.52 eV), for the $\alpha$-pinene in ice phase has lower intensity than that of $\beta$-pinene  which is also consistent with the gas phase and the theoretical calculations \cite{kubala2009}. Also, as observed in gas phase \cite{kubala2009}, the onset (excitation energy) in $\alpha$-pinene is much lower than $\beta$-pinene in ice phase. Comparing them with the onset in isoprene, though the $\alpha$-pinene, $\beta$-pinene and limonene are more complex than isoprene, the onset of absorption is much lower for these complex terpenes than that of isoprene. We also find a good agreement of our results with the VUV gas phase spectra of limonene, $\alpha$-pinene and $\beta$-pinene reported by Qiu et al., (2017)\cite{qiu2017}.

\section{Conclusions}

\par In this paper, we report the first VUV spectra of ices of isoprene and monoterpenes: limonene, $\alpha$-pinene and $\beta$-pinene, recorded at low temperatures (10 to 150 K). VUV photoabsorption of these molecular ices are quite similar to their gas phase counterparts. As an exception to this, isoprene absorption at the longer wavelength was found to be red shifted by nearly $\sim$20 nm; however, absorption intensity remained comparable. In the case of $\alpha$-pinene, the sudden reduction in a strong absorption band by warming the ice from 10 K to 30 K could be due to the temperature induced reorientation of $\alpha$-pinene molecules, leading to subtle changes in ice compaction. Of the molecules discussed in this work, isoprene, with a broad absorption band from 245 nm - 190 nm (5.06 eV - 6.52 eV) along with a distinct spectral shape and this absorption being far away from absorption spectrum of the dominant ice phase molecules such as water, oxygen, carbon dioxide and ammonia, seems to be a strong candidate for identification on icy bodies of the solar system. 

\section{Acknowledgements}
 We thank NSRRC for providing beamtime and the beamline facilities for the measurements. RR, JK, KKR, AB and BS acknowledges the support from PRL (Dept of Space, Govt of India).  RR, BS and NJM acknowledge the support from Sir John and Lady Mason academic trust. BS acknowledges the INSPIRE grant (IFA-11CH -11) for support to carry out part of this work during the period 2014 - 2017. AB was J.C. Bose Fellow during the period of this work. NJM acknowledges receipt of funding from the Europlanet 2024 RI the European Union’s Horizon 2020 research and innovation programme under grant agreement No 871149.

\bibliography{Isoprene_RR}

\end{document}